\documentclass[twocolumn,times]{aastex62}
\usepackage[utf8]{inputenc}
\usepackage[T1]{fontenc}
\usepackage[english]{babel}

\usepackage{savesym}
\savesymbol{tablenum}
\usepackage{siunitx}
\restoresymbol{SIX}{tablenum}

\usepackage{booktabs}
\usepackage[version=4]{mhchem}
\usepackage[capitalize,noabbrev]{cleveref}
\usepackage{txfonts}
\usepackage{csquotes}
\usepackage{textgreek}

\shorttitle{Revised Simulations of the PNLF}
\shortauthors{Valenzuela et al.}

\makeatletter
\usepackage{etoolbox}
\patchcmd\H@refstepcounter{\protected@edef}{\protected@xdef}{}{}
\makeatother

\usepackage[atend]{bookmark}
\bookmarksetup{
	numbered,
	addtohook={%
		\ifnum\bookmarkget{level}>1 %
		\bookmarksetup{numbered=false}%
		\fi
	},
}

\DeclareSIUnit{\solmass}{\mathnormal{M_\odot}} 
\DeclareSIUnit{\sollum}{\mathnormal{L_\odot}} 
\DeclareSIUnit\mag{mag}
\DeclareSIUnit\deg{deg}
\DeclareSIUnit\year{yr}
\DeclareSIUnit\parsec{pc}

\newcommand{\M}[1]{M\,#1}
\newcommand{\NGC}[1]{NGC\,#1}
\newcommand{\OIII}{[\ce{O}~III]}
\newcommand{\Hbeta}{\ensuremath{\ce{H}\beta}}
\newcommand{\Teff}{\ensuremath{T_\mathrm{eff}}}

\newcommand{\logT}{\ensuremath{\log\Teff}}
\newcommand{\logL}{\ensuremath{\log L}}
\newcommand{\logage}{\ensuremath{\log\mathrm{age}}}
\newcommand{\fit}{\ensuremath{r}}
\newcommand{\PNrate}{\ensuremath{\dot{\xi}}}
\newcommand{\samplesize}{\ensuremath{n_\mathrm{PN}}}
\newcommand{\Ltot}{\ensuremath{L_T}}
\newcommand{\lifetime}{\ensuremath{t_\mathrm{PN}}}

\newcommand{\zpoT}{\SI{81000}{\kelvin}}
\newcommand{\zpoL}{\num{2.75}}

\newcommand{\massMin}{.53}
\newcommand{\massMax}{.58}

\newcommand{\LtotNGC}{\SI{1.95e10}{\sollum}}
\newcommand{\PNrateNGC}{\SI{5e-12}{\per\year\per\sollum}}
\newcommand{\PNrateNGCold}{\SI{6e-12}{\per\year\per\sollum}}
\newcommand{\PNrateNGCerror}{\SI{2e-12}{\per\year\per\sollum}}
\newcommand{\LtotM}{\SI{8e10}{\sollum}}
\newcommand{\PNrateM}{\SI{1.2e-12}{\per\year\per\sollum}}
\newcommand{\PNrateMold}{\SI{1.7e-12}{\per\year\per\sollum}}

\begin{document}

\title{Revised Simulations of the Planetary Nebulae Luminosity Function}

\author{Lucas M. Valenzuela}
\affil{Fakultät für Physik, Ludwig-Maximilians-Universität, Schellingstraße 4, 80799 München, Germany}

\author{Roberto H. Méndez}
\affiliation{Institute for Astronomy, University of Hawaii, 2680 Woodlawn Drive, Honolulu, HI 96822, USA}

\author{Marcelo M. Miller Bertolami}
\affiliation{Instituto de Astrofísica de La Plata, UNLP-CONICET, Paseo del Bosque s/n, 1900 La Plata, Argentina}

\begin{abstract}

We describe a revised procedure for the numerical simulation of planetary nebulae luminosity functions (PNLF), improving on previous work (\citealt{mendez97}). The procedure now is based on new \ce{H}-burning post-AGB evolutionary tracks (\citealt{miller16}). For a given stellar mass, the new central stars are more luminous and evolve faster. We have slightly changed the distribution of the \OIII{} 5007 intensities relative to those of \Hbeta{} and the generation of absorbing factors, while still basing their numerical modeling on empirical information extracted from studies of galactic planetary nebulae (PNs) and their central stars. We argue that the assumption of PNs being completely optically thick to \ce{H}-ionizing photons leads to conflicts with observations and show that to account for optically thin PNs is necessary. We then use the new simulations to estimate a maximum final mass, clarifying its meaning,  and discuss the effect of internal dust extinction as a possible way of explaining the persistent discrepancy between PNLF and surface brightness fluctuation (SBF) distances.
By adjusting the range of minimum to maximum final mass, it is also possible to explain the observed variety of PNLF shapes at intermediate magnitudes. The new PN formation rates are calculated to be slightly lower than suggested by previous simulations based on older post-AGB evolutionary tracks.

\end{abstract}


\section{Introduction}
\label{sec:introduction}

If we can measure the nebular emission line fluxes of \OIII{} 5007 for many planetary nebulae (PNs) in a galaxy and transform the fluxes into apparent magnitudes $m(5007)$ (Jacoby magnitudes, defined by \citealt{jacoby89}), then we can build the planetary nebulae luminosity function (PNLF). It indicates how many PNs have \OIII{} 5007 at each apparent magnitude $m(5007)$. Empirically, it was discovered that the brightest PNs in a galaxy have similar absolute magnitudes $M(5007)$. This led to the suggestion of a universal bright end of the PNLF that could be used as a standard candle (\citealt{jacoby89}). The PNLF has been used to determine extragalactic distances for almost 30 years now (beginning with \citealt{jacoby89,ciardullo89}) and has proven to be a reliable distance indicator (see e.g. \citealt{ciardullo12}). 

The PNLF not only gives insight into the distance of a galaxy, but also into properties of the stellar population. Consider, for example, the PN populations of elliptical galaxies and of \M{31}'s bulge. The bright ends of their observed PNLFs require the existence of very luminous central stars, approaching \SI{7000}{\sollum} (see e.g. \citealt{jacoby99,mendez05,mendez08}). Because of the luminosity--core mass relation for post-AGB stars, this requires very massive central stars (about \SI{.63}{\solmass}), in strong disagreement with the expected maximum final mass of at most \SI{.55}{\solmass} that would correspond to a turn-off initial mass of about \SI{1}{\solmass} in these rather old stellar populations (\citealt{zhao12,cummings17,elbad18}). The problem has recently become more severe after individual values of internal dust extinction have become available for several of the brightest PNs in \M{31}'s bulge (\citealt{davis18}), suggesting some dereddened central star luminosities in excess of \SI{10000}{\sollum}.

This alarming discrepancy has been somewhat reduced by the introduction of new post-AGB evolutionary tracks by \citet{miller16}. A modified luminosity--core mass relation has decreased the required central star mass from \SIrange{.63}{.58}{\solmass} (\citealt{mendez17}), in better agreement with theoretical expectations. The new tracks also show a much faster post-AGB evolution than previously calculated. This means that central stars with much lower masses than previously expected can produce visible PNs.

If we further assume that the brightest PNs are predominantly optically thick (opaque) to \ce{H}-ionizing radiation from their central stars (\citealt{gesicki18}), it becomes possible to produce a bright end of the PNLF which stays almost invariant for a variety of stellar population ages and star formation histories. This helps to understand the high quality of the PNLF as a standard candle.

However, there are good reasons to doubt that most PNs are completely optically thick. Looking around in our Milky Way, we see PNs with predominantly bipolar symmetry, which means that they will probably start leaking \ce{H}-ionizing photons very soon, in directions where the nebular density is lower.

In view of all these complications, PNLF simulations created using Monte Carlo techniques can be used to take a closer look at how sample size, population age, and partial absorption of ionizing photons affect the properties of the PNLF (\citealt{mendez93,mendez97,mendez08}). The recent advances in post-AGB theory have induced us to revisit PNLF simulations and verify how far we are from really understanding the PNLF. 

In the present work we describe our revisions of the procedure outlined by \citet{mendez97} to generate PNLFs, using the new evolutionary tracks for low-mass stars (\citealt{miller16}) and a central star mass distribution derived from 
a modern DA white dwarf mass distribution (\citealt{kepler17}). The newly simulated PNLFs obtained in this way are compared with the observed PNLFs in a few selected galaxies to find out what the consequences are of the new tracks -- regarding the shape and bright end of the PNLF, the unresolved tension between PNLF distances and surface brightness fluctuation (SBF) distances (\citealt{ciardullo12,mendez17}), and the PN formation rate.

\section{Creation of evolutionary tracks}
\label{sec:create_lookup}

First of all, the reader might ask why our PNLF generation procedure follows the simple one described in \citet{mendez97}, instead of the more elaborate procedure described in \citet{mendez08}, which incorporated information derived from nebular models of \citet{SchIV07} as well as from observational work by \citet{Ciard02}. Let us justify this choice. Our 2008 paper included in its Figure 6 a comparison with Figure~2 of \citet{Ciard02}, showing good agreement. Furthermore, Figure~7 of \citet{mendez08} compared the PNLF bright end generated using the models of \citet{SchIV07} with the PNLF bright end produced with Mendez \& Soffner simulations, again showing good agreement. From these two comparisons we conclude that the method introduced in \citet{mendez97} is compatible with the empirical information later provided by \citet{Ciard02}. For that reason we avoided the approach used in the 2008 paper; it would have required a complete recalculation of the nebular model grid, using the new post-AGB evolutionary tracks of \citet{miller16}, characterized by different post-AGB central star masses and evolutionary speeds. We decided that for the purpose of studying the impact of the new evolutionary models on the PNLF it was simpler to use the 1997 simulation method.

It should be clear that a simulation based partly on random numbers cannot be perfect; the random numbers are indeed an attempt to compensate for the lack of complete information about nebular properties. For example, if we consider 5007 intensities, a full specification would require to know the PN metallicity distribution, about which we do not know enough. So our approach is to extract empirical information from observed cases. If we compare with theoretical nebular models like 
\citet{Dop92}, which predict that \enquote{at high metallicities up to \SI{15}{\percent} of the luminosity of the central star is reradiated in the 5007 line alone}, then we might find that we are generating a few PNs with a higher ratio L(5007)/Lstar, may be about \SI{18}{\percent}. However, any distorting effect produced by the generation of a few too high 5007/Lstar ratios through random numbers can be compensated by a few nebulae leaking too much ionizing radiation (which is expected according to Section~2.3 in \citealt{mendez08}), without affecting the general picture in any significant way. In summary, our simulations, although not perfect,
are good enough to study the effect of introducing the new post-AGB evolutionary tracks, which is our purpose here.

Since the procedure we have selected closely follows the one described in \citet{mendez97}, the reader may refer to that paper for more details. For brevity, here we will focus on the changes we have introduced. The evolutionary tracks of \ce{H}-burning post-AGB stars are now taken from \citet{miller16}. More specifically, we have selected five tracks corresponding to an overall metal mass fraction $Z_0 = \num{.01}$ from his Table 2, with initial masses \SIlist{1.0; 1.25; 1.5; 2.5; 3.0}{\solmass}. These tracks are based on improved models of AGB stars and evolutionary calculations. The main differences with previous work are that the new post-AGB stellar luminosities are higher and post-AGB timescales are shorter than what earlier models suggested. 

As before, we have chosen to define the post-AGB age to be equal to the time passed since the moment the central star had a temperature of $\Teff = \SI{25000}{\kelvin}$. We will call the post-AGB ages just \enquote{ages} in what follows. Because the central star mass changes continuously with time (see Eq.~(6) in \citealt{miller16}), each track is assigned the average mass between the masses at ages \numlist{0;30000} years. This hardly affects the outcome of the interpolated tracks because the change of mass along each track is much smaller than the central star mass itself, by an order of \numrange{e-4}{e-5}: mass loss rates for central stars inferred from observed spectra using the theory of radiatively driven stellar winds are of the order of \numrange{e-7}{e-9} $M_{\odot}$ yr$^{-1}$ (\citealt{kud2006}).

We generated a look-up table similar to the one described in \citet{mendez97} to determine \logT{} and \logL{} as functions of \num{3000} ages between \numlist{0;30000} years, and \num{260} masses between \numlist{.530;.706} solar masses. For the interpolations, the tracks were split up into three sections.

For the nearly horizontal heating tracks we used \num{40} temperatures between \SIlist{25000;81000}{\kelvin}, at which we plotted \logage{} and \logL{} as functions of mass. To interpolate between the values given by the tracks, we fitted curves to these plots and calculated the age and luminosity for the \num{260} masses at each of the \num{40} temperatures.

For the white dwarf cooling tracks we used \num{60} luminosities between $\logL = \num{1.60}$ and \zpoL, at which we plotted \logage{} and \logT{} as functions of mass. We fitted curves to these plots and calculated the age and temperature for the \num{260} masses at each of the \num{60} luminosities.

For the curved section joining the heating and cooling tracks we used \num{90} lines radiating at different angles from a fixed point at $\Teff = \zpoT$ and $\logL = \zpoL$. The values obtained at the intersections between these lines and the given tracks were used to plot \logage{}, \logT{} and \logL{} as functions of mass. We then fitted curves to these plots to calculate age, temperature and luminosity for the \num{260} masses at each of these \num{90} angles.

Having \num{190} values of temperature and luminosity for each of the \num{260} masses along their respective tracks, we could interpolate between these to obtain $\logT$ and $\logL$ at \num{3000} ages between \numlist{0;30000} years. The result was a look-up table that can be used to determine the temperature and luminosity of a star with a given age and mass by bivariate spline interpolation.

\begin{figure}
	\epsscale{1.17}
	\plotone{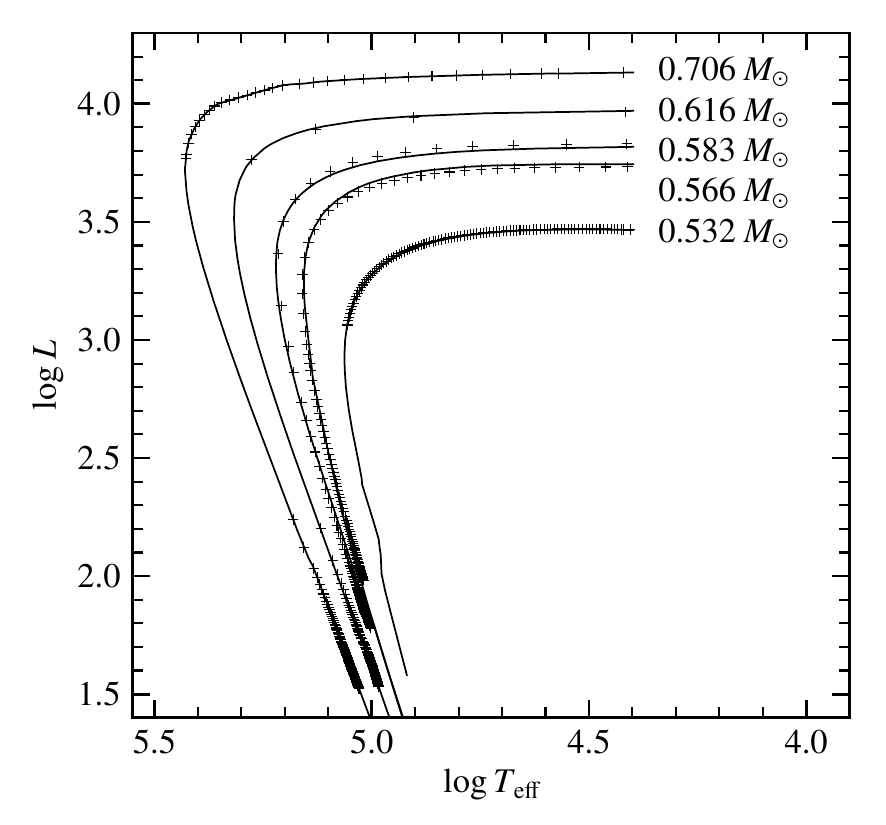}
	\caption{
		Solid lines show the five evolutionary tracks used to calculate our interpolations (\citealt{miller16}). Plus signs indicate the interpolated values of temperature and luminosity of central stars for the five masses at 100 evenly distributed ages between \numlist{0;30000} years. There are an additional \num{30} values for the \SI{.706}{\solmass} track at \num{30} evenly distributed ages between \numlist{0;300} years. When comparing this figure with Fig.~1 in \citet{mendez97}, note that the masses corresponding to a given luminosity are different, and that the post-AGB timescales of the new tracks are considerably shorter.
	}
	\label{fig:mass_tracks}
\end{figure}

\Cref{fig:mass_tracks} shows the evolutionary tracks for the five final (central star) masses used to calculate the interpolations.

\section{Generation of a PN population}
\label{sec:generate_sample}

When generating a sample of PN central stars, each star is assigned a random age and mass. The ages are uniformly distributed between \numlist{0;30000} years -- this is approximately the expected timescale of a PN, given typical observed expansion velocities and the largest observed PN sizes. 

\begin{figure}
	\epsscale{1.17}
	\plotone{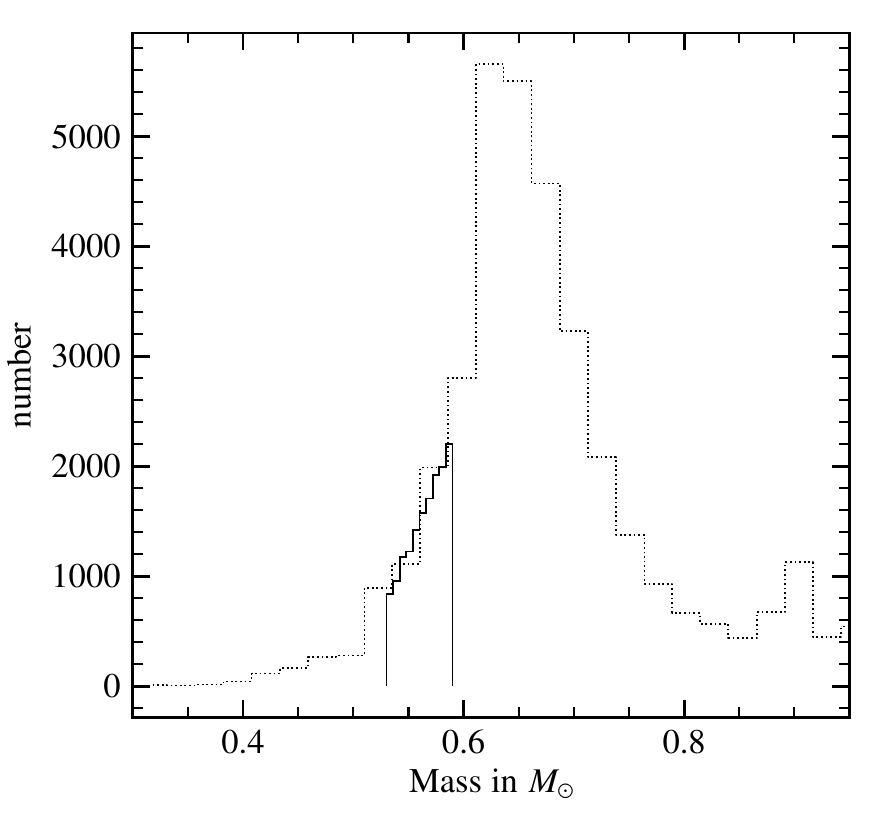}
	\caption{
		Histogram of \num{15000} PNs (solid line) with central star masses between \SIlist{.53;.59}{\solmass}, compared with the DA white dwarf mass distribution from \citet{kepler17} (dotted line). The latter was scaled to match the chosen sample size in this mass range.
	}
	\label{fig:mass_distribution}
\end{figure}

Because of the faster post-AGB evolution, central stars with lower masses than previously expected can produce visible PNs; but of course there must be a limit. PNs with central star masses below \SI{\massMin}{\solmass} should dissipate before the central stars can become hot enough to ionize the gas (see \cref{fig:mass_tracks}). Therefore, we generated masses between \SI{\massMin}{\solmass} and a maximum final mass that we expected to be somewhere below \SI{.60}{\solmass}. We approximated this range of masses as a linear distribution, with a mass of $\SI{\massMax}{\solmass}$ being \num{2.5} times as likely as a mass of $\SI{\massMin}{\solmass}$. This approximation is based on the DA white dwarf mass distribution for $\Teff \geq \SI{13000}{\kelvin}$ corrected by the $1/V_\mathrm{max}$ method in \citet{kepler17}. We compare both mass distributions in \cref{fig:mass_distribution}.

\begin{figure}
	\epsscale{1.17}
	\plotone{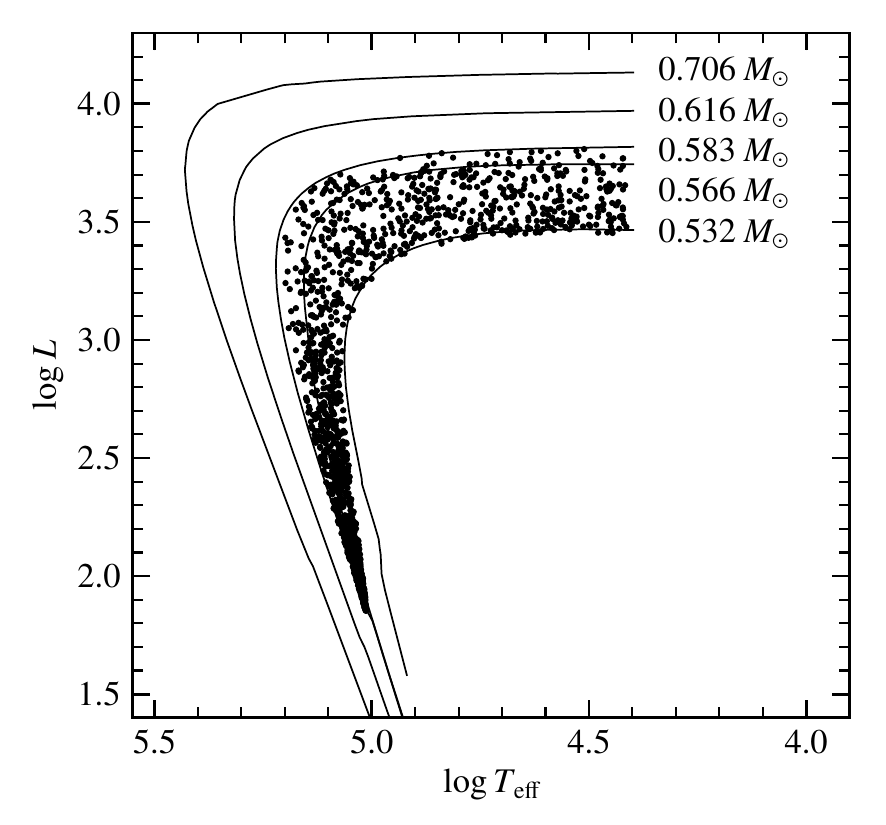}
	\caption{
		Solid lines show the same tracks for the five central star masses used in \cref{fig:mass_tracks}. Dots indicate the values of temperature and luminosity of \num{1500} central stars with random ages and masses. The ages are generated from a uniform distribution between \numlist{0;30000} years. The masses are generated using a linear distribution between \numlist{\massMin;\massMax} solar masses, with a mass of $\SI{\massMax}{\solmass}$ being \num{2.5} times as likely as a mass of $\SI{\massMin}{\solmass}$.
	}
	\label{fig:hr_sample}
\end{figure}

The HR diagram of \num{1500} PNs generated in this manner is shown in \cref{fig:hr_sample}. The values of central star temperature and luminosity were computed using the procedure described in \cref{sec:create_lookup}. 

For a PN which is completely optically thick to the Lyman continuum, it is possible to derive the \Hbeta \ luminosity for given values of central star temperature and luminosity under the assumption of a black body energy distribution (\citealt{ost2006}). See more details in \citet{mendez93}.

The simulated central star mass distribution was based as much as possible on empirical information. Unfortunately, for obvious reasons, the only such information available is for local white dwarfs. What would be the difference with populations without any recent star formation? We believe that the most important difference would be the presence in the local white dwarf sample of more massive white dwarfs, produced by those massive stars which have been able, despite their more recent birth, to complete their evolution and become white dwarfs.
But it is precisely those massive white dwarfs which we eliminate from our sample by introducing the \enquote{maximum final mass}. Therefore we do not expect any problem from this kind of age difference. On the other hand, we will discuss possible differences in the history of the star formation rate, e.g.\ the possible lack of lower-mass white dwarfs,
in \cref{sec:shape} below, in connection with the PNLF shape.

\section{Adjustments to the \texorpdfstring{\OIII{}}{[O III]} 5007 relative intensities}
\label{sec:intensity_ratios}

To compensate for the changes in mass distribution and evolutionary tracks, we slightly adjusted the procedure that generates the \textlambda5007 intensity, $I(5007)$, relative to the intensity of \Hbeta. We continued to compare our distribution to the one found in the LMC (using \num{118} PNs) and to the one found in the Milky Way (using \num{983} PNs). In both cases, we used the compiled data described in Section 4 of \citet{mendez97}.

\begin{figure}
	\epsscale{1.17}
	\plotone{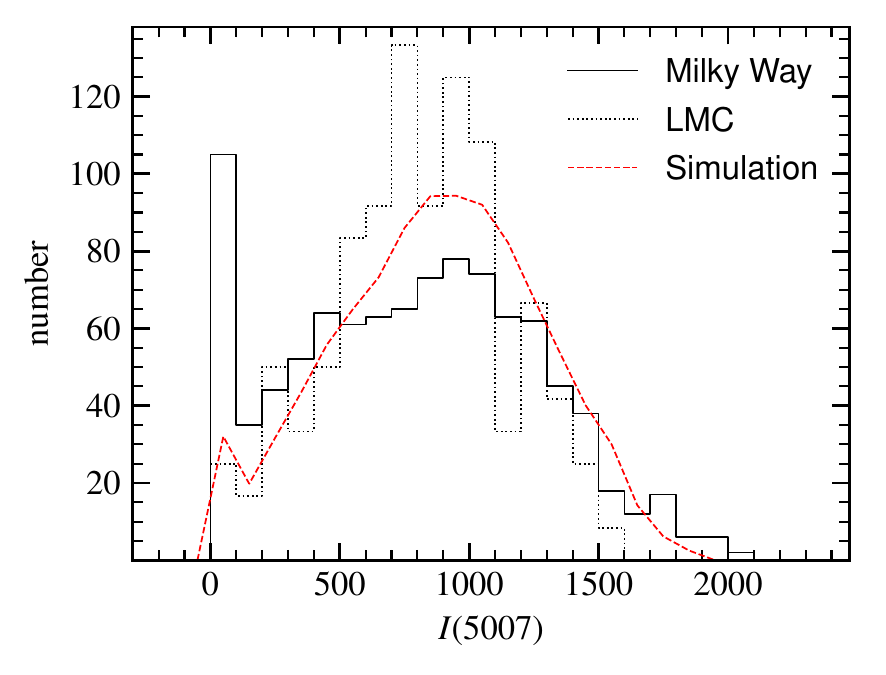}
	\caption{
		Our generated distribution of $I(5007)$ on a scale of $I(\Hbeta) = \num{100}$ compared with the observed distributions in the LMC (using 118 PNs) and the Milky Way (using 983 PNs). The histograms of our generated distribution and of the LMC have been normalized to the number of Milky Way objects.
}
	\label{fig:hist_ratio}
\end{figure}

Our distribution is generated by following the previous procedure, with the following differences: We now use a Gaussian distribution centered at $I(5007) = \num{950}$ on a scale of $I(\Hbeta) = \num{100}$ with $\mathrm{FWHM} = 375$. We then lower the values above \num{1200} by up to \num{200} (values right above \num{1200} are only slightly decreased while values around \num{2000} are maximally decreased) to imitate the steeper drop-off on the right part of the distribution. We then cap the values at \num{1850}. Because of the lower masses being used, we lower $I(5007)$ by $\SI{60}{\percent}$ for central stars with masses below $\SI{.55}{\solmass}$ that are on heating tracks with $\Teff > \SI{75000}{\kelvin}$ (instead of by $\SI{50}{\percent}$ for masses below $\SI{.57}{\solmass}$; following the discussion in Section 4 of \citealt{mendez93}). The new values were chosen to allow our generated distribution to fit the shapes of the observed ones in the LMC and the Milky Way as well as possible (see \cref{fig:hist_ratio}). We did not try to reproduce the Milky Way peak at very low $I(5007)$ because it has no effect on the bright end of the PNLF. 

For comments about the relative insensitivity of $I(5007)$ and its effect on the PNLF as a function of metallicity, please refer to \citet{jac1992}.

\section{The necessity of the absorbing factor \texorpdfstring{$\mu$}{\textmu}}
\label{sec:absorption_factor}

In \citet{mendez92} an absorbing factor $\mu$ was introduced to describe the optical thickness of a nebula in the \ce{H} Lyman continuum. Its value is equal to the fraction of ionizing stellar photons that are absorbed by the nebula, which means that $\mu = 1$ represents a completely optically thick or opaque PN, whereas a perfectly transparent PN has $\mu = 0$. A range of absorbing factors was used in PNLF simulations by \citet{mendez97}.

Now that we were using new post-AGB tracks and a different central star mass range, we had to make sure whether this absorbing factor was still required by the simulation to reproduce observed PNLFs. We assumed for the following that all PNs are optically thick, that is, $\mu = 1$ for all of them. Having the temperatures, luminosities, and intensities of \textlambda5007 relative to $I(\Hbeta)$, we could generate a PNLF following the procedure described in \citet{mendez97}.

For the observed data, we used a statistically complete PNLF with \num{119} PNs of \M{31}'s bulge (samples A and B from \citealt{ciardullo89}). We transformed apparent into absolute \textlambda5007 magnitudes adopting a distance modulus of $m - M = \num{24.43}$ and a foreground extinction correction for \textlambda5007 of \SI{.28}{\mag}, based on $E(B-V) = \num{.08}$ (\citealt{jac1992}.

At this point we introduced a quantity \fit{} that tells us how well a generated PNLF fits the observed one. We defined \fit{} to resemble the way the eye evaluates the fit in a plot. The sum of the squared deviations of $\log{N}$ seemed to be a good indicator, providing a more mathematical procedure than we used in previous PNLF papers. This fitting procedure can be described as chi-square minimization in log-space. We used a histogram for $M(5007)$ between \numlist{-6;-2} with a bin size of \num{.2} to match typical PNLFs found in the literature.
Since numbers below $\log{N} = 0$ are much less relevant and because of $\lim_{N\to0^+} \log{N} = -\infty$, we used $\log(\num{.5})$ instead of $\log{N}$ for values below $\log(\num{.5})$. We used this minimum value because $N < \num{.5}$ means that the probability of finding a PN in that particular bin is less than \SI{50}{\percent}.

\begin{figure}
	\epsscale{1.17}
	\plotone{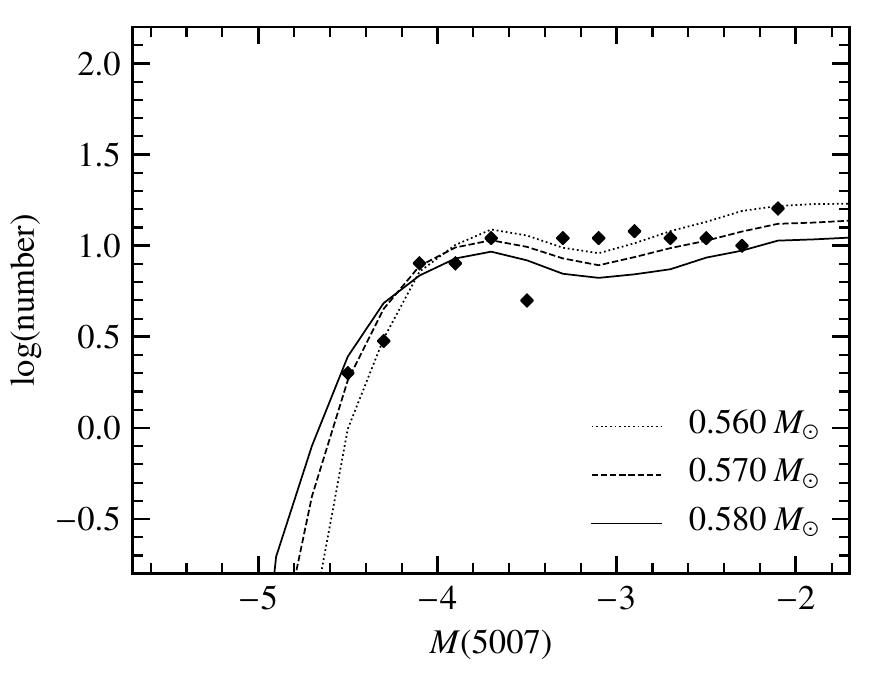}
	\caption{
		Statistically complete observed PNLF of \M{31}'s bulge (using 119 PNs) from \citet{ciardullo89}. We use a \textlambda5007 extinction correction of \SI{.28}{\mag} and a distance modulus of $m - M = \num{24.43}$. 
		The data are binned into \SI{.2}{\mag} intervals. The observed PNLF (diamonds) is compared with simulated PNLFs using completely opaque PNs ($\mu = 1$ for all of them). The sample size for each simulated PNLF is \num{285} PNs. The simulations were run with three different mass ranges from \SI{.530}{\solmass} to the respective maximum final mass displayed at the bottom right.
	}
	\label{fig:pnlf_M31_opaque}
\end{figure}

By adjusting the maximum final mass and the sample size in our simulated PNLFs, we found that a mass range of \numrange{.530}{.570} solar masses and a sample size of \num{285} PNs gives the best fit to the observed PNLF of \M{31} (see \cref{fig:pnlf_M31_opaque}).

Specifically, when calculating \fit{} for the three selected maximum final masses in \cref{fig:pnlf_M31_opaque}, we got \num{.30} for \SI{.560}{\solmass}, \num{.20} for \SI{.570}{\solmass}, and \num{.37} for \SI{.580}{\solmass}. Clearly, a maximum final mass of \num{.570} solar masses gives us the best fit.

While the bright end can be fitted properly, the shape of the PNLF beyond $M(5007) = -3.5$ has a valley that does not match the observed PNLF particularly well. And, most importantly, the sample size of \num{285} PNs is too small if we compare it to \num{970}, the total expected number of PNs in the region of \M{31}'s bulge sampled by \citet{ciardullo89}; see the discussion in their Section V.

If we now increase the sample size by a factor 3, while keeping \M{31}'s PNLF at the right distance, the simulated PNLF fails to fit. We conclude, as already stated by \citet{mendez97}, that it is necessary to allow for PNs that leak ionizing photons from the central star. The empirical evidence for this is taken from Table~4 in \citet{mendez92}, where we find a variety of $\mu$ values, from \num{1} to less than \num{.1}, and a clear trend of decreasing $\mu$ with increasing central star temperature. This UV photon leaking is further discussed by \citet{mendez08}.

\begin{figure}
	\epsscale{1.17}
	\plotone{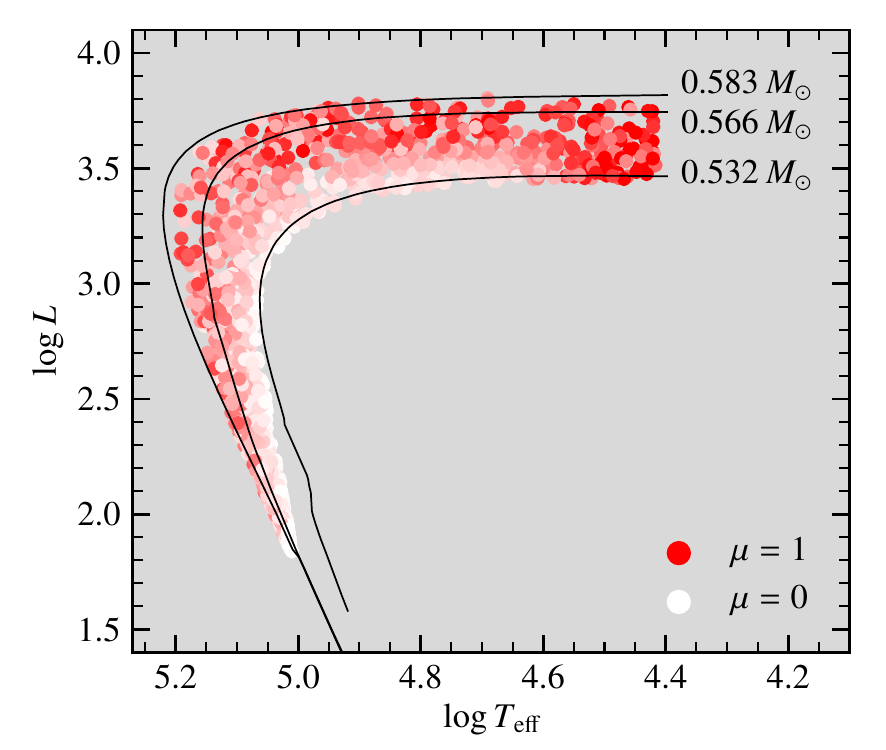}
	\caption{
		HR diagram showing the distribution of absorbing factor values in a sample of \num{1500} PNs that were generated with ages between \numlist{0;30000} years and masses between \numlist{\massMin;\massMax} solar masses. The color lightness of each PN represents its absorbing factor $\mu$, where red stands for $\mu = 1$ (optically thick) and white for $\mu = 0$ (completely transparent).
	}
	\label{fig:hr_sample_mu}
\end{figure}

We have decided to avoid using any theoretical guidance about nebular optical depth taken from modern nebular modeling efforts (e.g. \citealt{SchVII10}). Such radiation-hydrodynamics models are 1D, in other words spherically symmetric. The observed predominance of non-spherically symmetric PNs introduces uncertainty in our ability to predict what fraction of H-ionizing radiation is being lost through the nebular poles. Therefore, as in previous work, we generated a distribution of $\mu$ values using random numbers. Further justification for this approach can be found in Section 2.4 of \citet{mendez08}.
We implemented similar boundary conditions as \citet{mendez93} and \citet{mendez97}: PNs at the beginning of the heating tracks tend to be more opaque and will become increasingly transparent at higher temperatures. For the slowly evolving low-mass central stars (see \cref{fig:mass_tracks}), we expect the nebula to dissipate before reaching higher temperatures, so we assign low random absorbing factor values to them for higher ages. And finally, we use a similar method as \citet{mendez97} for generating decreasing $\mu$-values for increasing ages on the cooling tracks, such that $\mu = 0$ for an age of \num{30000} years. The resulting absorbing factor distribution in the HR diagram is shown in \cref{fig:hr_sample_mu}.

\section{Estimating the maximum final mass}
\label{sec:max_mass}

\begin{figure}
	\epsscale{1.17}
	\plotone{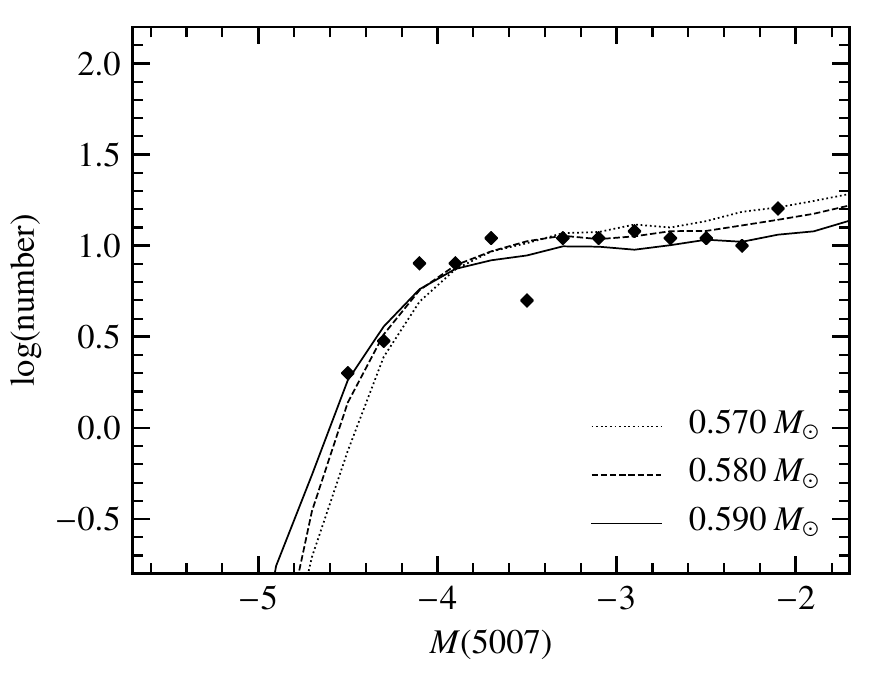}
	\caption{
		Statistically complete observed PNLF of \M{31}'s bulge (diamonds, same data as in \cref{fig:pnlf_M31_opaque}), compared with simulated PNLFs using the absorbing factor distribution from \cref{fig:hr_sample_mu}. The sample size for each simulated PNLF is \num{700} PNs. The simulations were run with three different mass ranges from \SI{.530}{\solmass} to the respective maximum final mass displayed at the bottom right.
	}
	\label{fig:pnlf_M31_masses}
\end{figure}

Using the absorbing factor distribution from the previous section, we could again generate PNLFs for different maximum final masses. The best fit to \M{31}'s observed PNLF was found at a sample size of \num{700} and a maximum final mass of between \num{.580} and \num{.590} solar masses (see \cref{fig:pnlf_M31_masses}). Again, we used the fitting parameter \fit{} defined in \cref{sec:absorption_factor}: We obtained \num{.39} for \SI{.570}{\solmass}, \num{.18} for \SI{.580}{\solmass}, and \num{.14} for \SI{.590}{\solmass}. This clearly gave us a better fit than when using completely opaque PNs ($\fit = \num{.20}$ for \SI{.570}{\solmass}; see \cref{fig:pnlf_M31_opaque}). We omitted masses beyond \SI{.590}{\solmass} since the slopes of their PNLFs' bright ends do not match the observed one. Because of the distribution of absorbing factors, we required a much larger sample size. This is more reasonable for \M{31}'s bulge and confirms that it is clearly preferable to allow for many optically thin PNs when simulating PNLFs.

At this point it is relevant to mention a recent paper by \citet{davis18} reporting PNs in \M{31} with high internal extinction and very luminous and massive central stars. Ideally, we would like to know all the individual extinction values and plot a fully dereddened PNLF. Since we only know a few individual extinction values, that is currently not possible. In fact, \citet{davis18} show that correcting just a few PNs destroys the typical PNLF shape (see their Figure~14). The only practical option is to assume an average value for the extinction. If we adopt a higher average extinction, the whole PNLF shifts towards higher luminosities, and therefore the maximum final mass needs to increase. We will come back to this point later on.

Because of the small number of PNs surveyed by \citet{ciardullo89} in \M{31}'s bulge, the uncertainty of the maximum final mass is high. For a better estimate of this value, we needed to consider an old stellar population with a larger sample size. For this purpose we used \NGC{4697}, a flattened elliptical galaxy in the Virgo southern extension. For the observed data, we used the method described in \citet{mendez01} to get a statistically complete bright end of the PNLF consisting of \num{328} PNs. The data were taken from the corresponding catalog (\citealt{mendez08cat}).

On the assumption that the PNLFs of \M{31}'s bulge and \NGC{4697} are identical, we adopted the distance modulus $m - M = \num{30.1}$ determined by the PNLF method (\citealt{mendez01}). The foreground extinction correction for \textlambda5007 was \SI{.105}{\mag}, based on $E(B-V) = \num{.03}$ from \citet{tonry01}.

\begin{figure}
	\epsscale{1.17}
	\plotone{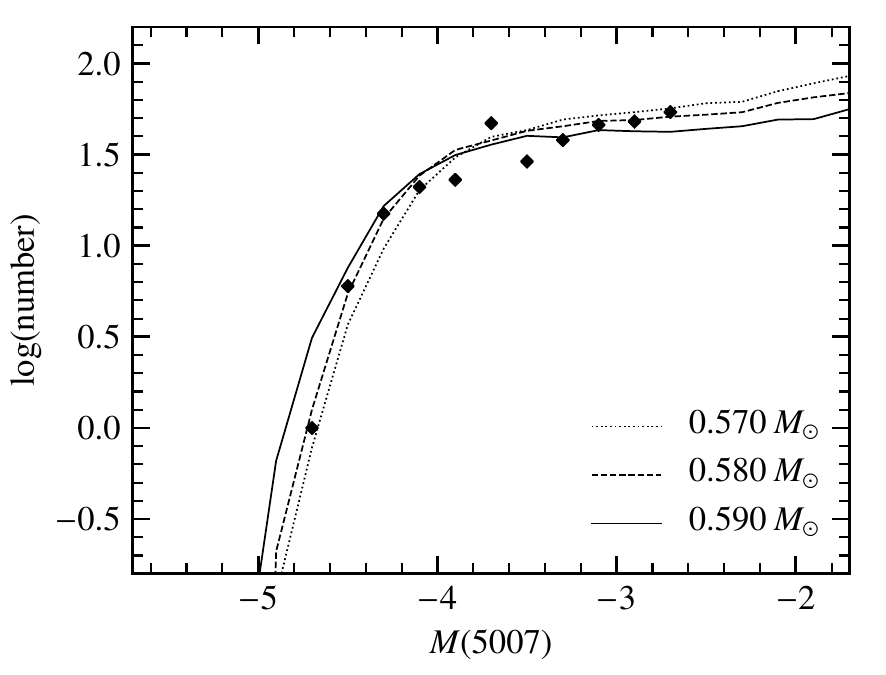}
	\caption{
		Statistically complete observed PNLF of \NGC{4697} (328 PNs) from \citet{mendez01}. We use a \textlambda5007 extinction correction of \SI{.105}{\mag} and a distance modulus of $m - M = \num{30.1}$. 
		The data are binned into \SI{.2}{\mag} intervals. The observed PNLF (diamonds) is compared with simulated PNLFs using the absorbing factor distribution from \cref{fig:hr_sample_mu}.  The sample size for each simulated PNLF is \num{3000} PNs. The simulations were run with three different mass ranges, from \SI{.530}{\solmass} to the respective maximum final mass displayed at the bottom right.
	}
	\label{fig:pnlf_NGC4697_masses}
\end{figure}

Again using the absorbing factor distribution from the previous section, we fitted our simulated PNLF to the observed one. We found the following best-fitting values: a mass range of \numrange{.530}{.580} solar masses and a sample size of \num{3000} PNs (see \cref{fig:pnlf_NGC4697_masses}). Calculating the fit factor \fit{} for the three maximum final masses gave us \num{.16} for \SI{.570}{\solmass}, \num{.09} for \SI{.580}{\solmass}, and \num{.34} for \SI{.590}{\solmass}. This time, we got a very clear best fit. The reason for this is the much larger sample size in \NGC{4697} compared to that in \M{31}. We conclude that our new PNLF simulations work very well with a maximum final mass of \SI{.58}{\solmass}, confirming a preliminary result by \citet{mendez17}.

To analyze the properties of PNs at the bright end of the simulated PNLF with a maximum final mass of \massMax{}, we took a closer look at the PNs with absolute \textlambda5007 magnitudes brighter than \SI{-4.2}{\mag}. In our simulations we found that the distribution of absorbing factors for these PNs has extreme values of \num{1.0} and \num{.56}, a mean value of \num{.89}, and a standard deviation of \num{.10}.

We would like to emphasize that our PNLF simulations do not include internal dust extinction.
The empirical existence of a \enquote{maximum final mass} does not mean that more luminous and massive central stars
cannot exist. We need to add the condition that if more luminous central stars do exist (as in the bulge of \M{31}), they are always affected by internal dust extinction. This is not hard to imagine, because we can expect the more massive central stars to eject more material and to evolve more quickly, leading necessarily to a denser distribution of circumstellar material. The consequence would be that the PNs originating from the more massive central stars would pile up at or near the bright end of the PNLF, mixed with those PNs produced by central stars with maximum final mass, no extra dust extinction, and absorbing factors close to \num{1}. Of course this interpretation will need to be tested with individual spectroscopic studies of complete samples of extragalactic PNs defining the bright end of the PNLF. 

\section{Interpretation of SBF distances in terms of the maximum final mass}
\label{sec:discussion_mass}

The use of the PNLF for distance determinations is based on the assumption that the PNLF's bright end is universal. Any systematic difference between PNLF distances and those determined by other methods is a challenge to that assumption. 
The SBF method of distance determination (\citealt{tonry88, tonry01, blakeslee09}) has shown a tendency to produce larger distance moduli than the PNLF method, by about \SIrange{.3}{.4}{\mag} (\citealt{teod10}). We refer the reader to \citet{ciardullo12} for a discussion of the most likely cause for this systematic effect: zero point errors arising from limited knowledge of internal extinction in the calibrator galaxies, to which the PNLF and SBF methods react in opposite ways. In what follows we would like to illustrate the consequences of assuming that the SBF distances of elliptical galaxies are correct.  

\begin{figure}
	\epsscale{1.17}
	\plotone{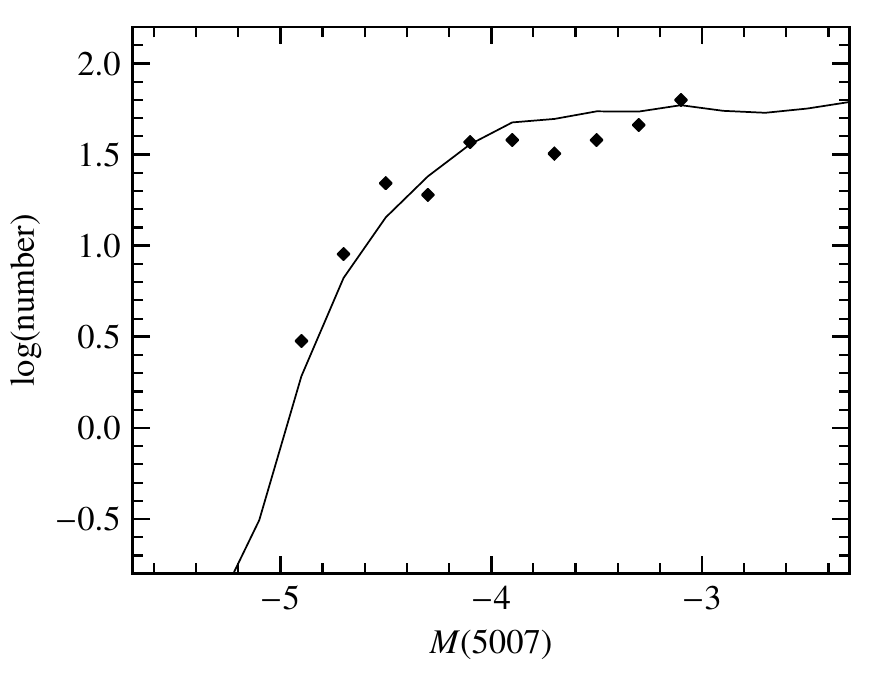}
	\caption{
		Statistically complete observed PNLF of \NGC{4697} (diamonds, 328 PNs) using the same extinction correction \SI{.105}{\mag} as before, and an SBF distance modulus of $m - M = \num{30.4}$ instead of the PNLF distance modulus. In order to fit our simulated PNLF (solid line), we need a maximum final mass of \SI{.61}{\solmass} and a sample size of \num{5200}. The absorbing factor distribution is, as before, from \cref{fig:hr_sample_mu}.
	}
	\label{fig:pnlf_NGC4697_SBF}
\end{figure}

An increased distance forces the PNLF to become more luminous. What maximum final mass would be required for our simulated PNLF to fit an observed PNLF if shifted to the SBF distance? Consider for example \NGC{4697}.
\citet{blakeslee09} used the SBF method to determine a distance modulus $m - M = \num{30.491\pm.065}$. We adopted an SBF distance as close to the PNLF distance as allowed by the SBF error bars: $m - M = \num{30.4}$. As in \cref{sec:max_mass}, we used an extinction correction of \SI{.105}{\mag}. With those numbers we transformed apparent into absolute \textlambda5007 magnitudes. In order to get a fit with our simulated PNLF we needed to increase the maximum final mass to \num{.61} $M_{\odot}$. The sample size also increased to \num{5200} (see \cref{fig:pnlf_NGC4697_SBF}). The fit parameter was $\fit = \num{.34}$.

\begin{figure}
	\epsscale{1.17}
	\plotone{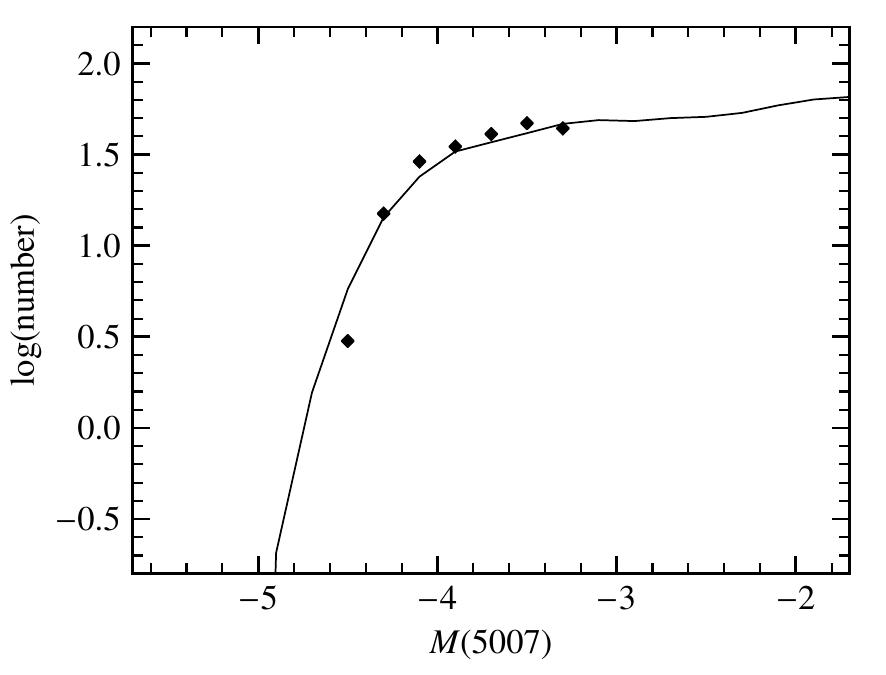}
	\caption{
		Statistically complete observed PNLF of \M{60} (diamonds, 218 PNs) using an extinction correction of \SI{.09}{\mag} and the PNLF distance modulus $m - M = \num{30.7}$, compared to our simulated PNLF (solid line) with a maximum final mass of \SI{.58}{\solmass}, a sample size of \num{2950}, and the absorbing factor distribution from \cref{fig:hr_sample_mu}. 
	}
	\label{fig:pnlf_M60}
\end{figure}

As another example we took \M{60}, an elliptical galaxy in the Virgo Cluster. For the observed data, we used the method described by \citet{teodorescu11} to get a statistically complete sample of \num{218} PNs. We adopted their PNLF distance modulus of $m - M = \num{30.7}$ and extinction factor of \num{.09} in our calculations. After fitting the simulated PNLF to the observed one, we got a sample size of \num{2950} and a maximum final mass of \SI{.58}{\solmass} (see \cref{fig:pnlf_M60}). The fit parameter was $\fit = \num{.26}$.

\begin{figure}
	\epsscale{1.17}
	\plotone{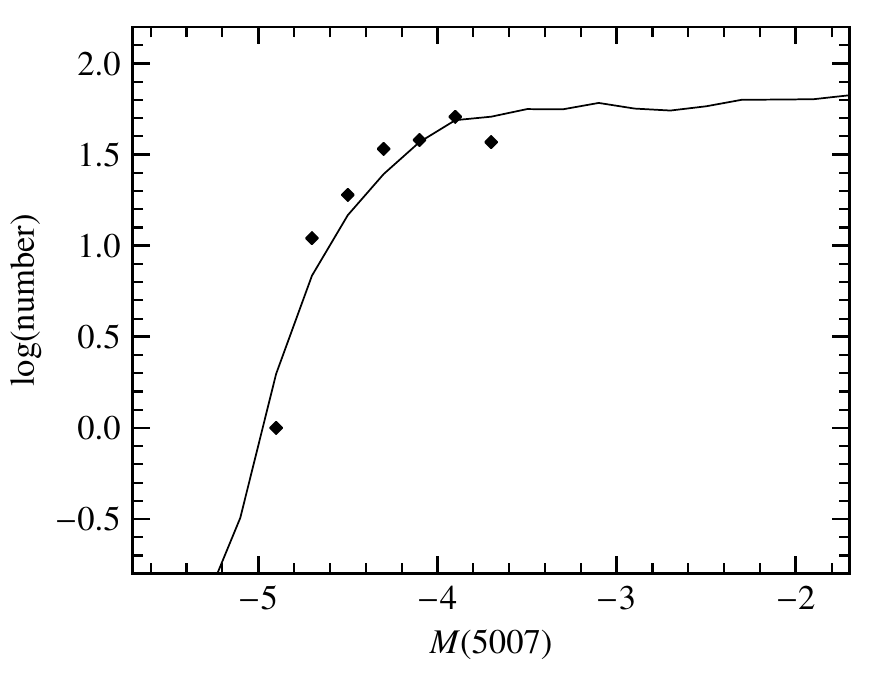}
	\caption{
		Statistically complete observed PNLF of \M{60} (diamonds, 218 PNs) using the same extinction correction \SI{.09}{\mag} as before, and an SBF distance modulus of $m - M = \num{31.0}$ instead of the PNLF distance modulus. In order to fit our simulated PNLF (solid line), we need a maximum final mass of \SI{.61}{\solmass} and a sample size of \num{5350}. The absorbing factor distribution is, as before, from \cref{fig:hr_sample_mu}.
	}
	\label{fig:pnlf_M60_SBF}
\end{figure}

Next, we adopted an SBF distance as close to the PNLF distance as allowed by the SBF error bars from \citet{blakeslee09}: $m - M = \num{31.0}$ (they found $m - M = \num{31.082\pm.079}$), and we recalculated the absolute \textlambda5007 magnitudes. Again, to force a fit with our simulated PNLF, we needed a maximum final mass of \SI{.61}{\solmass} and a sample size of \num{5350} (see \cref{fig:pnlf_M60_SBF}). The fit parameter was $\fit{} = \num{.28}$.

\begin{table}
	\caption{Values used for PNLF and SBF distances of \NGC{4697} and \M{60}, and the respective best fitting parameters of the simulation. The fit parameter \fit{} is defined in \cref{sec:absorption_factor}.}
	\begin{tabular*}{\columnwidth}{@{\extracolsep{\fill}}lcccc}
		\toprule
		& \multicolumn{2}{c}{\NGC{4697}} & \multicolumn{2}{c}{\M{60}} \\
		\cmidrule(rl){2-3} \cmidrule(rl){4-5}
		& PNLF & SBF & PNLF & SBF \\
		\midrule
		distance modulus & \num{30.1} & \num{30.4} & \num{30.7} & \num{31.0} \\
		extinction correction & \num{.105} & \num{.105} & \num{.09} & \num{.09} \\
		sample size & \num{3000} & \num{5200} & \num{2950} & \num{5350} \\
		maximum final mass & \num{.58} & \num{.61} & \num{.58} & \num{.61} \\
		fit parameter & \num{.09} & \num{.34} & \num{.26} & \num{.28} \\
		\bottomrule
	\end{tabular*}
	\label{tab:pnlf_sbf_parameters}
\end{table}

A summary of our tests is given in \cref{tab:pnlf_sbf_parameters}.
In both cases (\NGC{4697} and \M{60}) the PNLF distance leads to a maximum final mass of \SI{.58}{\solmass}, whereas the SBF distance leads to \SI{.61}{\solmass}. The problem with accepting the SBF distances is, first, that we do not expect such massive central stars to originate from an old stellar population if we assume single post-AGB stellar evolution; and second, that the PNs defining the bright end of the PNLF in galaxies like \NGC{4697} and \M{60} become more luminous than any PNs ever discovered in our Local Group. If SBF distances are confirmed, we will need to find explanations for these anomalies.

Now we see that \citet{davis18} may have provided such an explanation. Imagine that \M{31}'s bulge (perhaps every spiral bulge) has PNs that suffer from higher average values of internal dust extinction than what we find in elliptical galaxies like \NGC{4697} and \M{60}, perhaps as a consequence of a higher average metallicity. Assuming a higher average extinction for \M{31} than previously adopted, we need to shift its PNLF towards higher luminosities. In order to recover the fit, since we are using \M{31} as a calibrating galaxy, we need to shift the elliptical galaxy's PNLF by increasing its distance modulus to values closer to those obtained with the SBF method. Note that the difference between the average circumstellar extinctions for \M{31} and \NGC{4697} reported by \citet{davis18} is \SI{.37}{\mag}, similar to the systematic difference between SBF and PNLF distance moduli for distant elliptical galaxies. 

There would be a price to pay: all the galaxies then would have very bright PNs, requiring very luminous and massive central stars.  We would agree with \citet{davis18} that this situation would require rethinking the whole interpretation of the PNLF, probably requiring a significant contribution from other factors than only single post-AGB star evolution.

The impact would not be limited to PN research; our current understanding of stellar populations comes from population synthesis models, which assume in particular that we understand AGB stars. A careful revision of every result obtained using the old AGB and post-AGB stellar evolution calculations would then become a very prudent precaution.

Probably the best way to test SBF and PNLF distances will be to use Tip of the Red Giant Branch (TRGB) distances. The method is described by \citet{makarov06}; it has been applied to many galaxies up to distances of about \SI{10}{\mega\parsec}. Ideally, we need at least ten TRGB distances in the Virgo and Fornax clusters to yield a statistically convincing result. This will be possible with the James Webb Space Telescope (JWST), which is expected to allow TRGB distance determinations up to at least \SI{30}{\mega\parsec}. If the SBF distances are confirmed to be closer to the truth than PNLF distances, then the explanation in terms of different amounts of average internal extinction becomes very likely.

Further investigation of this idea (different average internal dust extinction) would probably be best oriented toward obtaining additional deep spectrograms of PNs in elliptical galaxies like \M{60}, with the purpose of measuring the individual Balmer decrements. Metallicity determinations from integrated light analysis indicate that NGC 4697 may be more metal-poor than the bulge of M31, while \M{60} (NGC 4649) may be not (see \citealt{Trager00}). But this may not be decisive, because of the presence of a metallicity gradient (with detected PNs located predominantly in low surface brightness regions) and because of unknown complications in the details of dust formation.

\section{PN formation rates}
\label{sec:formation_rate}

A fit to the observed PNLF with our simulated PNLF provides a simultaneous determination of distance modulus and sample size (the total number of PNs in the surveyed area of the galaxy). What is the effect of the new \citeauthor{miller16} tracks on PN formation rates?
Having obtained new sample sizes, we can recalculate the specific PN formation rate \PNrate{} with
\begin{equation}
	\samplesize = \PNrate \Ltot \lifetime
\end{equation}
where \samplesize{} is the sample size, \Ltot{} is the total luminosity of the sampled population, and $\lifetime = \SI{30000}{\year}$ is the lifetime of a PN.
We take \Ltot{} from \citet{mendez01} for \NGC{4697}: $\Ltot = \LtotNGC$. With the new sample size $\samplesize = \num{3000}$ from \cref{sec:max_mass} we get $\PNrate = \PNrateNGC$. This is slightly lower than the earlier result, $\PNrate = \PNrateNGCold$ (\citealt{mendez01}). It should be noted that the uncertainty continues to be high at around ${\pm \PNrateNGCerror}$.

We can perform the same calculation for \M{60}. The total bolometric luminosity of the sample from \citet{teodorescu11}, $\Ltot = \LtotM$, and the determined sample size $\samplesize = \num{2950}$ from \cref{sec:discussion_mass} are used to get $\PNrate = \PNrateM$. Again, this is lower than the earlier result, $\PNrate = \PNrateMold$ (\citealt{teodorescu11}). The uncertainty is similarly high for \M{60}.

Since post-AGB stellar evolution is much faster along the \citeauthor{miller16} tracks than what previous models suggested, one might have expected a higher PN formation rate in order to provide the amount of PNs we observe. However, the final mass range has also changed. Since we are using lower final masses, the central stars move more slowly along their respective tracks in the HR diagram. Ultimately, this leads to a somewhat smaller PN formation rate, just as we can see in our calculations for \NGC{4697} and \M{60}. Consequently the new tracks do not help to solve the old discrepancy between stellar death rates of about 15 $\times 10^{-12}$ yr$^{-1} L_{\odot}^{-1}$ (\citealt{renBuz86}) and PN formation rates a factor 3 (or more) smaller. We still need to assume that most pre-white-dwarf stars in galaxies like NGC 4697 and \M{60} cannot produce a visible PN, or can at most produce a short-lived one (\citealt{mendez93}).

\section{PNLF shapes}
\label{sec:shape}

\citet{mendez97} found a change of slope between magnitudes $M(5007) = \num{-3.5}$ and \num{-2.3}. While our simulations do not entirely eliminate this feature, it is much less noticeable since we no longer have any negative slope in that interval. The PNs in this section of the PNLF are mostly on heating tracks and in the curved section joining the heating and cooling tracks. Comparing our distribution of PNs in the HR diagram (\cref{fig:hr_sample}) with the one in Figure~2 of \citet{mendez97}, it is clear that we now have a much larger amount of PNs with luminosities right below $\logL = \num{3.5}$ in the heating tracks and the curved section. The distribution of PNs in this region of the HR diagram is also more uniform. Therefore, it seems that the lack of such PNs in \citet{mendez97} had been the reason for the change of slope in the PNLF. This can be traced to the existence of a quick drop in luminosity as the \ce{H}-burning shell is extinguished and the central star enters the white dwarf cooling track. 

It is important to note that all the \ce{H}-burning central stars show the luminosity drop, but it is more pronounced for the more massive ones (see \cref{fig:mass_tracks}). Thus the adoption of the \citet{miller16} tracks and substantially lower central star masses leads to predicting a monotonically increasing PNLF for an old population, which seems to be confirmed by \NGC{4697}. However, other galaxies, like the LMC and SMC, do show a \enquote{camel-shaped} PNLF (\citealt{jacoby02, reid10}). This kind of shape might be explained by involving slightly more massive central stars, but also by a relative lack of lower-mass stars (see Section~5 of \citealt{mendez08}). We can test this by applying our PNLF simulations to the LMC.

\citet{reid10} present a sample of PNs in the LMC over a large range of magnitudes. We used their sample of \num{577} PNs in the central \SI{25}{\deg\squared} region of the LMC (Tables~1 and 2 in the corresponding catalog, \citealt{reid10cat}). For the transformation into absolute \textlambda5007 magnitudes, we adopted their calculated distance modulus of \num{18.50} and an extinction correction of \SI{.46}{\mag}, based on $E(B-V) = \num{.13}$ (\citealt{massey95, soffner96}).

Instead of the observed LMC PNLF, corrected with an average foreground extinction, we could have taken the PNLF with each PN corrected for individual internal extinction, also given by \citet{reid10}; but both LMC PNLFs, shown by Reid \& Parker in their Figures 6 and 7, show very similar camel shapes. Therefore for a discussion of PNLF shapes it should not matter which one is used for the comparison.

\begin{figure}
	\epsscale{1.17}
	\plotone{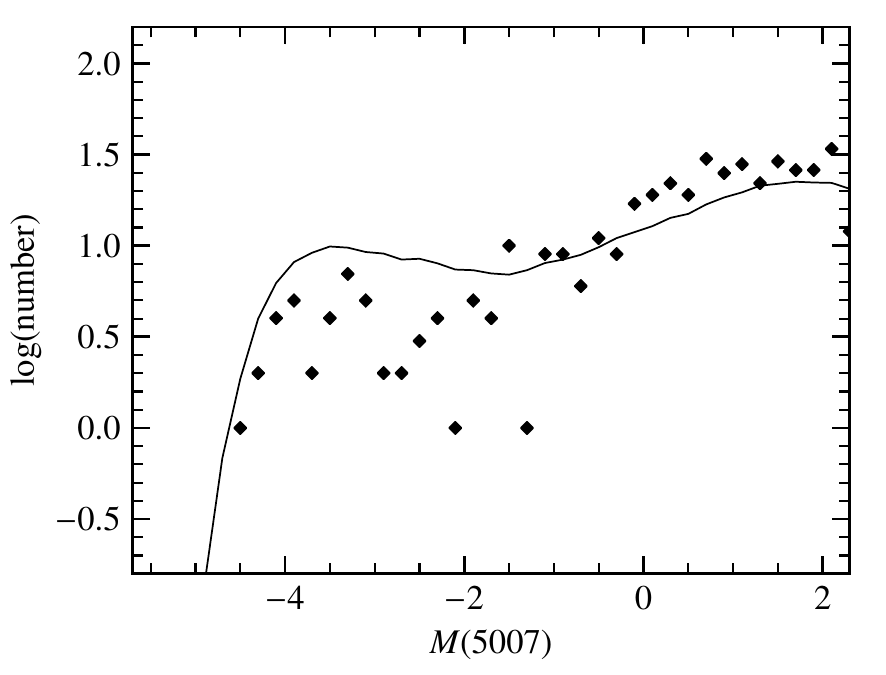}
	\caption{
		Observed PNLF of the central \SI{25}{\deg\squared} region of the LMC (\num{577} PNs) from \citet{reid10cat}. We use a \textlambda5007 extinction correction of \SI{.46}{\mag} and a distance modulus of $m - M = \num{18.50}$.
		The data are binned into \SI{.2}{\mag} intervals. The observed PNLF (diamonds) is compared with our simulated PNLF (solid line) with a modified central star mass range (\SIrange{.55}{.59}{\solmass}). The absorbing factor distribution is from \cref{fig:hr_sample_mu}. The sample size for the simulated PNLF is \num{600} PNs.
	}
	\label{fig:pnlf_LMC}
\end{figure}

Fitting the simulated PNLF to the observed one, we found that we can obtain the valley in the PNLF between $M(5007) = \num{-3.5}$ and \num{-.5} (see \cref{fig:pnlf_LMC}), similar to what we can see in \citet{reid10}. As expected from our earlier discussion, we had to increase both the minimum and maximum final masses to \SIlist{.55;.59} to get this \enquote{camel shape}. Since there are complicating factors in this case (possible differences in the age of the population and in extinction corrections), we do not expect a perfect agreement, but are satisfied that the \enquote{camel shape} in the LMC can be reproduced by changing just a few parameters.  We note in passing that \citet{badenes15} have reported that the LMC PNs show a limited range of ages. The lack of a very old stellar population would seem to agree with the lack of very low mass central stars we need to reproduce the camel shape.

It should be clear that we are far from knowing all the details that contribute to generate the PNLF in any stellar population. To mention just a few, we have the effects of binary evolution and the existence of a minor (but significant) fraction of \ce{He}-burning central stars in addition to the \ce{H}-burners. Such complications are very hard to model in Monte Carlo style. To all this we need to add the possibility of different amounts of internal dust extinction, which affects the maximum final mass required to fit the PNLF. The full potential of the PNLF for studies of stellar populations will probably come only after careful testing of PNLF distances and more individual studies of bright PNs in different galaxies, which will likely require \num{30}-meter telescopes.

\section{Conclusion}
\label{conclusion}

We have explored the influence of the new post-AGB evolutionary tracks by \citeauthor{miller16} on the interpretation of the observed PNLFs in other galaxies. For that purpose we have revised an earlier procedure for numerical PNLF simulations (\citealt{mendez97}). Some adjustments were made to several routines without altering the basic philosophy of the method, which is to rely on empirical information about nebular and stellar properties. A comparison of our simulations with the observed PNLF of \M{31}'s bulge, whose distance is very well known, shows that we need to allow for optically thin PNs. Having established the distribution of absorbing factors $\mu$ with \M{31}, we selected a galaxy with a larger sample size, \NGC{4697}, to get a better estimate for the maximum final mass. We clarify the meaning of the maximum final mass by remarking that it does not preclude the presence of more luminous and massive central stars within PNs suffering internal dust extinction. We briefly illustrate the consequences of adopting SBF distances as a way of remarking on the importance of solving the discrepancy between PNLF and SBF distances. We also report on the effect of the new evolutionary tracks on the calculation of PN formation rates. Finally, we show that by adjusting the minimum final mass that leads to a visible PN, we can reproduce the \enquote{camel shape} exhibited by PNLFs in the Magellanic Clouds.

\section*{Acknowledgments}
\label{acknowledgments}

L.M.V. acknowledges partial support from the German Academic Scholarship Foundation and would like to thank the Institute for Astronomy, University of Hawaii at Manoa, for their kind hospitality during the course of this research. We thank the anonymous referee for many useful comments.

\bibliography{bib}{}
\bibliographystyle{aasjournal}

\end{document}